# Comparison of caloric effects in view of application

Zhongjian Xie[1], Gael Sebald, Daniel Guyomar

Université de Lyon, INSA-Lyon, LGEF EA682, 8 rue de la Physique, 69621 Villeurbanne, France

## Abstract

In the framework of solid-state cooling technology, four kinds of caloric effects, magnetocaloric (MC), electrocaloric (EC), barocaloric (BC) and elastocaloric (eC) effects, are compared in view of environment discussion and application fields. This field is primarily developed in the intention of protecting the environment. However, some widely researched caloric materials do not meet environmental friendly criteria. Currently, new caloric materials (called "good" material) possessing the properties of friendly environment, low cost, high caloric performance and practicability (low stress) need to be found. In view of application, all current caloric effects/materials are investigated for the common objective of cooling. However, the adapted application cases for different caloric effects/materials are not considered. Due to the different stimuli and different intrinsic properties (different caloric performance for different dimensions) of caloric materials, they can exhibit unique advantage to different application cases. "Good" caloric material used for large-scale application is still less and deserve more search. Purposeful research will save the cost and accelerate the commercialization.

## 1. Introduction

**Motivation**

Environmental problem is becoming more and more prominent. It mainly comes from the development of industrial technology. The new technology only pursuing the high performance is not sustainable and acceptable. It is important to use the environmental standards as the primary criterion for technological development.

Nowadays, the refrigeration market is large and keeps growing. Vapor-compression has been the dominant technology for all cooling applications during the past 100 years. Although its cooling performance is improving, its basis on hazardous gas has not changed. This has resulted in many environmental problems. Besides the ozone depletion, the problem of global warming becomes prominent. The global warming potential (GWP) of the conventional refrigerants (hazardous gases) used in vapor-compression equipment is as high as 1000-2000 times that of $CO_2$[1]. Even though we are aware of the catastrophic global warming, the global mean surface temperature increased continuously up to 0.7 K from 1960 due to human activity[2]. According to the latest report by Deconto and Pollard[3] in the journal Nature, "Antarctica has the potential to contribute more than a metre of sea-level rise by 2100 and more than 15 metres by 2500, if the greenhouse gas mission continue unabated". Hence, we should keep environment standards in mind and it is urgent to develop more sustainable cooling techniques.

---

[1] zhongjian.xie@insa-lyon.fr
gael.sebald@insa-lyon.fr

**Basic trend of caloric effect**

Solid-state cooling technology based on caloric effects is drawing more and more attention. The caloric effects mainly include the magnetocaloric (MC), electrocaloric (EC), barocaloric (BC) and elastocaloric (eC) effects. They refer to the isothermal entropy change or adiabatic temperature change upon application or removal of magnetic field, electrical field, hydrostatic pressure and uniaxial stress, respectively. BC and eC effects are referred as mechanocaloric (mC) effect. The caloric effect was first found by Gough in natural rubber (NR) and further investigated by Joule[4]. The MC effect and materials have already been studied for a long time[5–7]. The EC effect has been studied more seriously from recently[8–10]. The mC effect is the least studied[4,11–13]. However, due to the large temperature change (around 20 K) of eC effect of shape memory alloys (SMAs)[14,15], the trend of their application is emerging recently. It appears some patents[16], some simulation papers of cooling device[17–19] and the fabrication of prototype based on the eC effect of SMAs[20]. According to a report by the US Department of Energy (2014)[1], eC cooling technology shows the largest potential among all the alternatives to vapor-compression technologies.

**Consideration for environment and application**

The solid-state cooling system is driven by the needs of environmentally friendly technology. However, in the current research on caloric effects, many environmental problems appear. For MC effect, the needed magnets and the MC material are mainly based on the rare-earth elements (REEs), whose production is detrimental to environment[21]. The arsenic (As)-based MC material[7] and the lead-based EC materials possess high caloric performance but they are toxic. These materials show the largest potential for application and thus the largest potential for environmental harm. Some other caloric materials are environmentally friendly and show a high caloric performance but with a high cost, like the PVDF-based polymers[22]. The eC and BC effect of shape memory alloys (SMAs) need a large stress (several hundreds of MPa)[11,13,20,23–27], which is not practicable[17]. Thus, new caloric materials possessing the properties of friendly environment, low cost, high caloric performance and practicability need to be found.

The extensive descriptions of caloric effects have been existed in several reviews (for example refs [7], [10], [24] and [4]). In this review, the caloric effects for solid-state cooling are evaluated in view of environment and application. The eC effect of natural rubber (NR) is proved to be potential for cooling application and compared with other caloric effects. The different application cases of different caloric effects/materials are highlighted. Potential work for materials and devices is also proposed.

## 2. Magnetocaloric (MC) effect and its environmental discussion

MC effect has drawn large research interest. Its development has been well described in some review papers[28–30]. There appears over 40 prototype cooling devices but without commercialization. The major obstacle is the expensive magnets for strong magnetic field, which is also difficult to generate. The rare-earth elements (REEs) based permanent magnets (Nd–Fe–B) is likely to be employed as the magnetic field source. Since 0.5–10 kg of Nd will be used in one cooling device, the cost of the device will mainly depend on the cost of neodymium (Nd)[31]. Moreover, MC material is mainly based on REEs (Gd, $Gd_5Ge_2Si_2$, $La(Fe,Si)_{13}$ ). Although the MC cooling technology is regarded as an environmentally friendly technology, the production of REEs itself is far from environmentally sustainability as it requires significant material and energy consumption while generating large amounts of waste[21].

China is currently the largest producer of REEs[32], supply more than 90% of the world market. It has resulted in serious environmental concerns, with the heavy metal and radioactive emissions in groundwater, soil and plants[33] and flue dust containing HF, $SO_2$, and $H_2SO_4$[34]. In 2008, China began to limit the production of the REE and number of permitted exporters[21,35]. It has caused the global REEs supply shortages. In 2011 and 2012, the U.S. Department of Energy (DOE) released the Critical Materials Strategies, calling for a reduced dependence on critical materials, to ensure that related technologies are not obstructed by future supply shortages[36]. The most critical materials identified by the DOE are all REEs: dysprosium (Dy), terbium (Tb), europium (Eu), neodymium (Nd) and yttrium (Y)[37]. Thus, besides the environmental harm of the production of REEs, the MC cooling is not commercially competitive.

Different MC materials are compared under a field change of 2 T in Fig. 1[7]. It is obvious that, near room temperature, a few transition metal-based alloys perform the best[7]. However, they all contain arsenic (As) element. If the MC research continues for just pursuing good performance, the cooling system using these As-contained alloys may be commercialized. However, if it is released to the environment, the groundwater might be polluted. This environmental pollution will not be inferior to the hazardous gases from vapor compression.

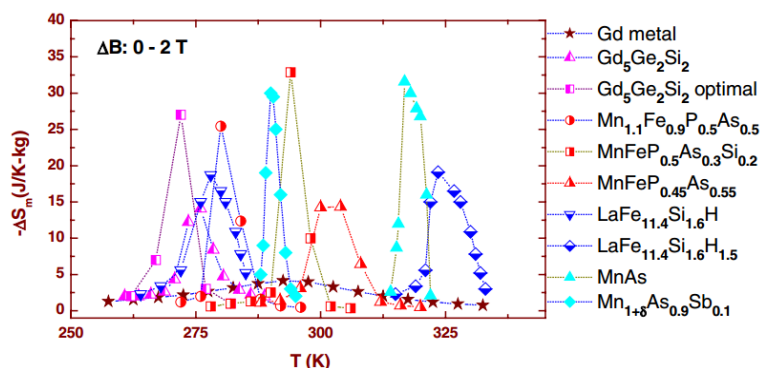

Fig. 1 Entropy change of different MC materials under magnetic field changes of 2 T [7].

Thus, for an environmentally friendly and commercially competitive MC cooling technology, firstly, the permanent magnet based on REEs cannot be used. Secondly, the MC materials based on REEs and As cannot be used. However, the small magnetic field of non-REE permanent magnet cannot meet the needs for stimulating the current MC materials. The environmentally friendly MC materials have a low performance. To solve both the problems of magnets and MC material, the multi-caloric effect may be employed. The hydrostatic pressure is mostly used to couple with MC effect[38–40]. The height, the width and the position of the peak in entropy change $|\Delta S(T)|$ can be modified by the pressure[4].

Currently, considering the permanent magnets and main MC materials based on REEs, and other toxic MC materials, MC cooling may not be regarded as an environmentally friendly technology. Thus, the responsibility of finding a real environmentally friendly technology falls on the EC effect and mC effect.

## 3. Electrocaloric effect and its adapted application

For EC effect, the history, experimental method and performance of different kinds of EC materials have been described in review[10]. This part focuses on the environmentally friendly EC materials and its adapted application.

Nowadays, most of the current high performance EC bulk materials contain a large amount of lead in contradiction to the original intention of the development of solid-state cooling technology.

The research on EC materials should focus on lead-free compounds. $BaTiO_3$ is an environmentally friendly EC material. PVDF-based polymers have also been investigated as a potential alternative[8] (see section 5.1).

$BaTiO_3$ is found to have a giant EC strength ($\Delta T/\Delta E=220$ mK.cm.kV$^{-1}$)[41]. It can also present the conventional eC[42] and inverse BC effect[43]. For EC and eC effect, the position of EC temperature change ($\Delta T$) peak is at around 400 K, which cannot be used for near-room temperature cooling application. However, for BC effect, the position of $\Delta T$ peak decreases as pressure increases. When a 2 GPa is applied, the position of $\Delta T$ peak is close to the room temperature. Since $BaTiO_3$ possesses these three caloric effects, it indicates that the polarization change of $BaTiO_3$ can be stimulated by electric field, uniaxial stress and hydrostatic pressure. Accordingly, a coupling can be produced between the three stimuli. The uniaxial stress shifts the EC peak to higher temperature[42]. Interestingly, the applied pressure can shift the EC peak to room temperature[43], which makes $BaTiO_3$ potential for near room temperature cooling application.

As the thin film EC material (oxide and polymer) has high caloric performance[10], the device study of EC material is based on the thin film. A cheap commercial Y5V $BaTiO_3$-based multilayer capacitors (MLC) has been demonstrated by Kar-Narayan and Mathur[44,45]. Based on their calculation, a house-hold air conditioner can be constructed from around 100,000 MLCs. It may be too complex and high costly for large-scale application. Actually, another driving force for the solid-state cooling technology is the small-scale application, where the vapor-compression technology with a low efficiency cannot be employed. Thus, the thin film EC material is more adaptive for the small-scale application. This is a good correspondence of the intrinsic property of material (high performance of thin film) with the application case (small volume and small cooling capacity of thin film for small-scale application).

## 4. Mechanocaloric (mC) effect of shape memory alloys (SMAs)

For mC effect, two kinds of materials are studied: one is the hard materials, shape memory alloys (SMAs) and the other is the soft materials, elastomers. SMAs are the mostly researched mC materials. Among them, Ni-Ti alloy (nitinol) is most promising. It exhibits a large eC $\Delta T$ of around 20 K and $\Delta S$ of 37 $J \cdot K^{-1} \cdot kg^{-1}$ (240 $kJ \cdot K^{-1} \cdot m^{-3}$)[14,15]. The BC effect of Ni-Mn-In shows a relatively low $\Delta T$ (4.5 K) (compared with eC effect)[23]. The comparison of different mC materials can be found in ref[24].

In the following part, we focus on the application challenges of hard materials (SMAs).

### 4,1 Application challenge of elastocaloric effect of hard materials

The active magnetic regenerator (AMR) has been applied in all up-to-date magnetic refrigerators, since it is the most efficient way to utilize the MC effect so far[29]. Such a refrigerator system by using eC effect of SMAs is proposed by Tusek *et al.*[17]. The system configuration and its four operational steps are shown in Fig. 2. It consists of parallel eC plates, two heat exchangers, a heat-transfer fluid, a fluid pumping system, and an actuator to (un)load the material.

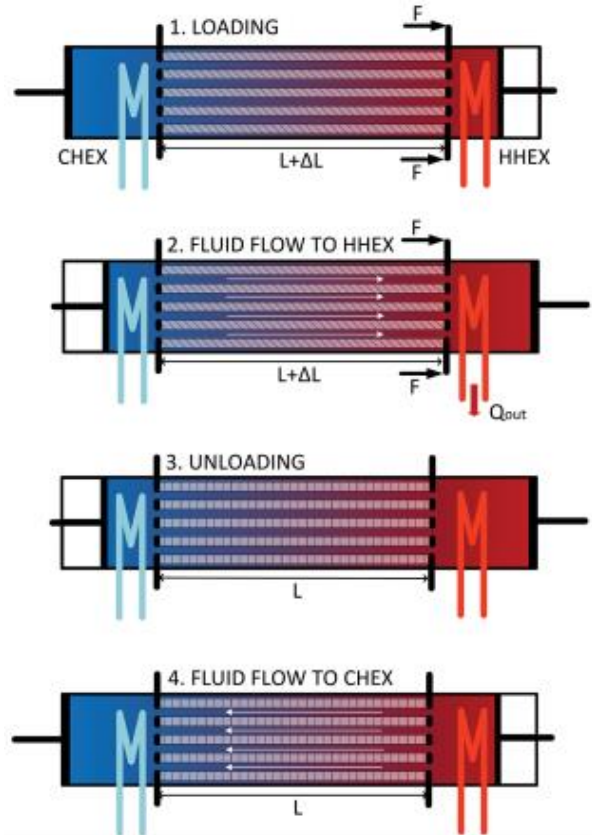

Fig. 2 Schematic presentation of the operation of the eC regenerator[17].

Another similar AMR system by using SMAs is proposed by Qian *et al.*[18]. Fig. 3 shows the schematic of a single SMA bed assembly using nitinol tubes. These tubes are used to produce cooling and heating capacity. Their holders are designed to sustain radial direction stress during their compression process, as well as to avoid bending. The two loading heads were originally designed to feed heat transfer fluid (HTF) into the nitinol tubes for heat transfer and transfer the compression force directly into the nitinol tubes.

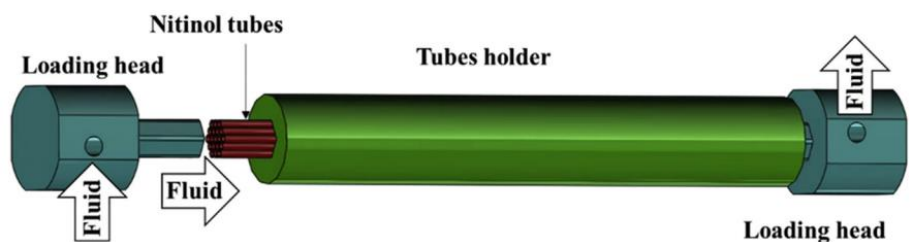

Fig. 3 Original design of the nitinol tubes bed for cooling[18].

One main drawback for the practical application of mC effect of SMAs (including eC and BC effect) is the large forces needed to perform the deformation. In Tusek's model, for 500 W of the cooling power, the evaluated system requires an applied force of about 180 kN (stress of 900 MPa) for Ni-Ti alloy and about 50 kN (stress of 275 MPa) for Cu-Zn-Al. To the knowledge of the authors, the commercialized linear motor can only provide a maximum force of several kN. Moreover, the large force will easily cause the damage of clamp or slip of material after a number of cycles. Consequently, the large force will decrease the system stability and increase the cost.

The problem of SMAs concerning the large force has already been highlighted by a real setup (Fig. 4)[20]. Schmidt *et al.* used two linear motors to output a force of 2 kN. This large force only allows for loading SMA (NiTi) with a small cross section of 1.6 mm$^2$. With a length of 90 mm, the cooling power is only 9 J in one cycle. It is difficult to use the bulk SMA material for the large-scale application. It is the similar situation to the bulk EC material which cannot be used for the cooling application due to the low dielectric strength[10]. Thus, the SMAs film may be only used. Their surface-to-volume ratio is higher than the bulk one, which is advantageous for fast heat transfer and allows for higher cycling frequencies. Due to the small volume of SMAs thin film and its small cooling capacity, it is interesting for small-scale system.

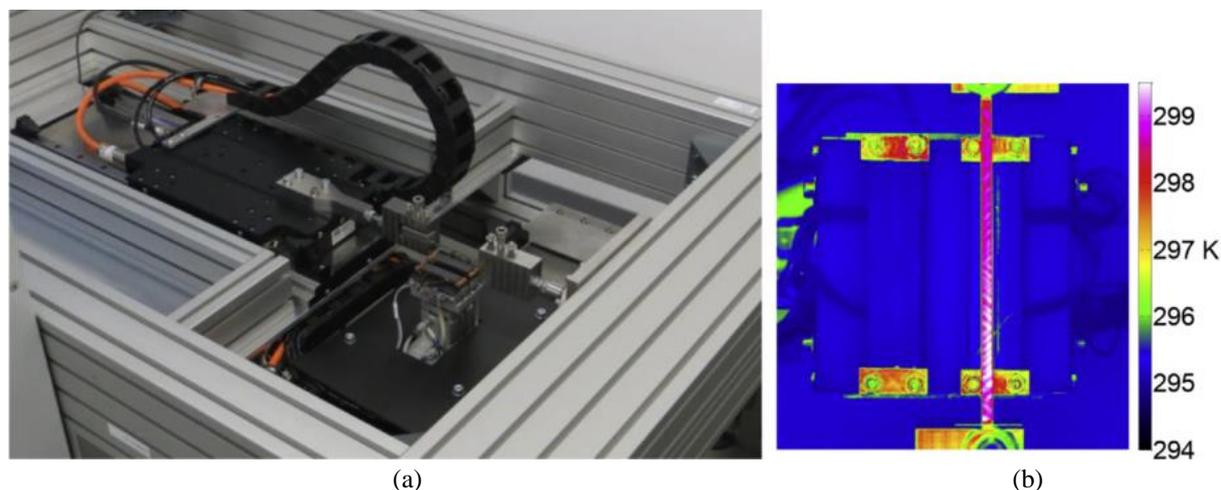

(a)          (b)

Fig. 4 (a) Cooling device based on eC effect of a NiTi ribbon. (b)An infrared thermal image reveals the temperature change established by repeatedly using the SMA NiTi ribbon to absorb heat at the heat source and then dump it to the heat sink[20].

Both current EC technology and eC technology based on thin films are potential for the small-scale application. Electric field seems easier to be provided than the stress for small-scale system, like IT device. A voltage generator is easier integrated in small-scale system than a force generator.

**4,2 Potential barocaloric direction**

All the current mC cooling devices are based on the eC effect of SMAs. The devices based on the BC effect have not yet been reported. Uniaxial stress is easier to be applied than a hydrostatic pressure, which needs to be applied in all axes. For eC effect, the positions for the application of uniaxial stress and for the heat exchange can be separate, which facilitates the design of the device. For solid BC material, the application position of hydrostatic pressure conflicts with the position of the heat transfer. This may be difficult for designing a cooling device based on solid BC material.

BC material can be fabricated into the form of particles and dispersed in liquid. A liquid BC material can then be produced. There are three advantages of the liquid BC material. Firstly, the problem of position conflict of heat transfer and pressure application in solid BC material can be solved. The pressure can be just applied in one axis. Then, all the liquid material will get a same hydrostatic pressure. The area for heat exchange can be in transverse axis. Secondly, there is no fatigue problem of the liquid BC material, whereas it is really a problem for the solid eC material. Thirdly, the current vapor-compression technology can facilitate the BC cooling technology transformation based on liquid BC material thank to the same driving mode.

## 5. Mechanocaloric effect of soft material (polymer)

Since the large hydrostatic pressure/uniaxial stress is the main drawback of mC effect of hard materials (SMAs), the mC effect of soft materials need to be investigated. Moreover, the compressor in current refrigerator normally works at around 1 MPa[46]. This pressure level has already produced the noise. Larger pressure/stress is not desired. Furthermore, in order to facilitate the technology transfer, the pressure/stress needed for mC material should not be too far from this stress level.

### 5.1 Caloric effect of PVDF-based polymers

For hard material, the BC effect comes from the transformation of the crystallographic structure[30,40]. For the polymers with pure amorphous structure, the BC effect may not occur. It may occur in some semi-crystalline polymers.

The giant EC effect of P(VDF-TrFE) and P(VDF-TrFE-CFE) was found by Neese *et al.* in 2008[22]. Following this finding, the EC effect of P(VDF-TrFE-CFE) was studied extensively[47]. However for the EC effect of P(VDF-TrFE-CTFE), only one publication reported a EC ΔT of 1.1 K by applying 50 MV/m[48]. The mC effect of PVDF based materials was just reported by Patel *et al.*[49]. For PVDF, a peak eC ΔT of 1.8 K was observed by applying 15 MPa at 298K. For P(VDF-TrFE-CTFE), a large BC ΔT of 6K was observed by applying 200 MPa at 300K. A larger BC ΔT of 19 K for the same pressure level occurred near 368 K. As the EC performance of P(VDF-TrFE-CFE) is better than P(VDF-TrFE-CTFE), what about the their BC effects?

The phase transition of P(VDF-TrFE-CTFE) responsible for BC effect is the α to β phase transition[50]. The β-phase is the most highly polar phase, whose unit cell consists of two all-*trans* chains packed with their dipoles pointing in the same direction[51]. The monomer CFE is more effective than CTFE in reducing the all-*trans* conformation[52]. Thus, the P(VDF-TrFE-CFE) may get a higher BC effect.

### 5.2 Elastocaloric effect of elastomer

Natural rubber (NR) is a natural material and environmentally friendly. The eC effect of NR was found by the natural philosopher John Gough in early nineteenth century. The same effects were reported for various materials by Joule in 1859[4]. Nowadays, there is relatively little research work about eC effect of NR in view of cooling application[53–55]. Even in the recent review papers[4,56–58]. NR has not been considered for eC application or compared with other caloric materials. The research of eC effect of NR is just beginning.

All elastomers can act as eC material due to the contribution of the alignment of polymer chains. For the elastomers exhibiting strain-induced crystallization (SIC), a giant eC effect may be obtained from the latent heat of phase transition. NR possesses SIC phase transition and is potential for a giant eC effect[59–61]. Elasticity (reversibility) and adiabatic temperature change are two required quantities for eC effect. Elasticity is the primary condition for eC effect. Both of them are mainly related to the SIC[61–64]. Due to the developed wide-angle x-ray diffraction (WAXD) by using synchrotron radiation[65], the SIC research is blossoming. It can help to understand the eC effect deeply and promote its cooling application.

#### 5.2.1 Mechanism of elastocaloric effect of natural rubber

Some research work on NR is related to the eC effect but in view of material science. Dart[66] got a maximum temperature change of ~12 K in NR (Fig. 5). The upturn of temperature change at elongation of 3 and hysteresis of temperature change in extension and retraction were observed. It indicates the occurrence of SIC and its dominant effect on temperature change. The temperature change of NR has been used to study the SIC kinetics[61,62]. In turn, SIC research can guide the eC research[54,55].

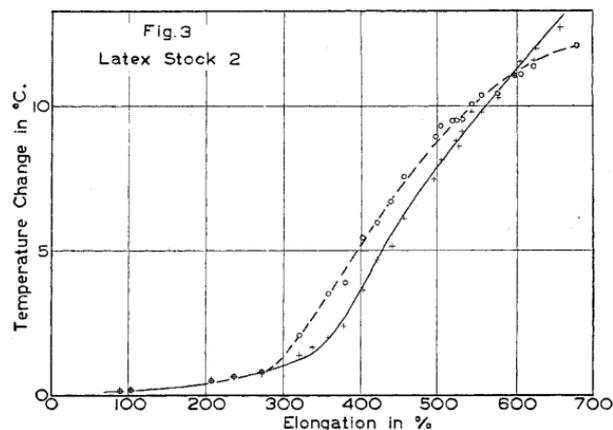

Fig. 5 Temperature changes in adiabatic extension (solid line) and retraction (dotted line)[66].

The crystal structure and the lattice parameters of SIC in NR can be deduced from the WAXD pattern. Fig. 6 gives a typical diffraction pattern that can be obtained when X-ray beam is perpendicular to the stretching direction of NR[67].

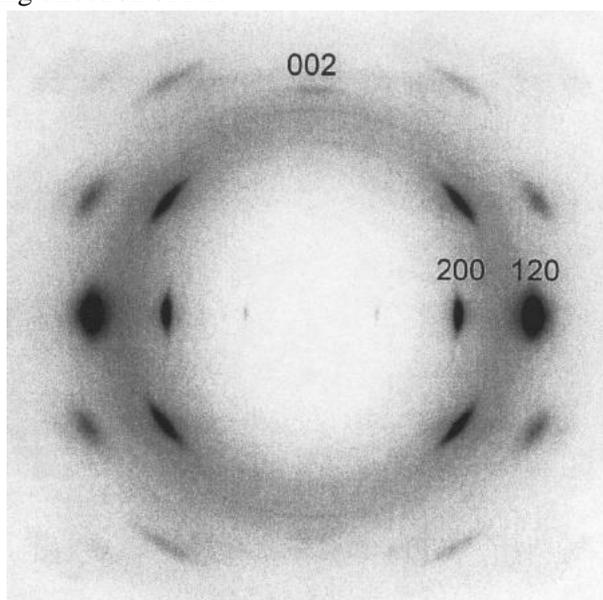

Fig. 6 Typical WAXD pattern of a vulcanized NR. The stretching direction is vertical[67].

The crystallinity-strain and stress-strain behaviors are shown in Fig. 7[68]. Like the temperature change-strain behavior (Fig. 5), they also exhibit hysteresis. The loop direction of crystallinity-strain curve is same with temperature change-strain curve (anticlockwise). It indicates that the crystallinity is responsible for the temperature change. However for stress-strain curve, the loop direction is opposite to those of temperature change-strain and crystallinity-strain curves. It's attributed to the complex stress effects (stress relaxation[69,70] and stress hardening[63,71]) of SIC, which is firstly proposed by Flory[72].

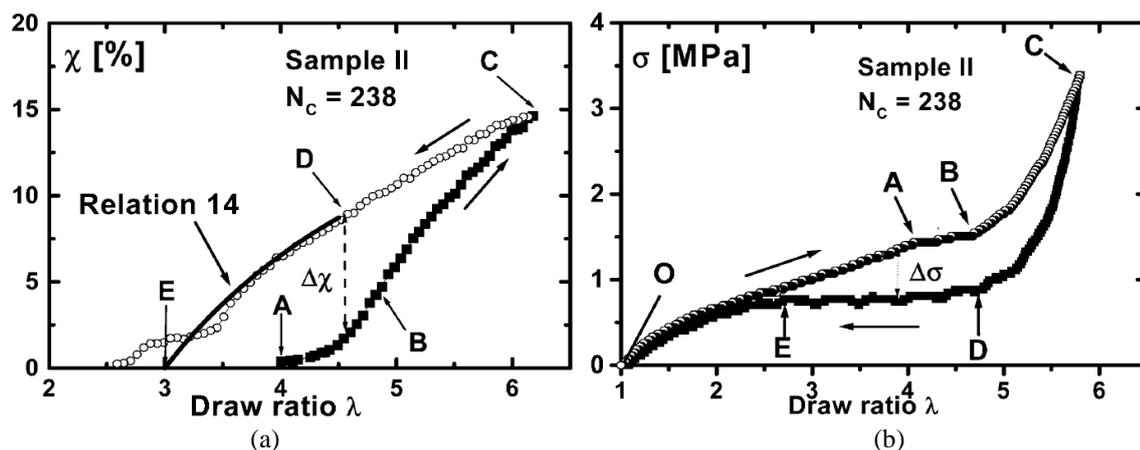

Fig. 7 Simultaneous evolutions of: (a) crystallinity and (b) stress vs elongation during a deformation cycle (strain rate ~ $5 \times 10^{-4}$ s$^{-1}$) at 22 °C for a vulcanized NR[68].

**5.2.2 The study on elastocaloric effect of natural rubber**

In Guyomar's work (2013)[53], the NR was considered seriously for eC cooling application and its refrigerant capacity was analyzed. Considering the design of the compact cooling device, the method of pre-strain was proposed for reducing the applied strain of NR[73]. It has been extensively evidenced that the SIC of NR occurs at around strain of 3 [68,71,74,75]. By applying a pre-strain, the amorphous strain regime (before the onset strain of SIC) can be skipped and the SIC strain regime with a large latent heat can be reached directly. Thus, a larger eC strength of NR represented by eC coefficient $\beta = -\partial s / \partial \varepsilon$ was obtained, where $s$ is the entropy and $\varepsilon$ is the strain. The fatigue life of eC material should be more important for the eC cooling applications. For the same strain amplitude of 3, the eC properties of amorphous strain regime (strain of 0-3) and SIC strain regime (strain of 2-5) were compared (in press). The fatigue life at amorphous strain regime is only about 2000 cycles. For the SIC strain regime, there is no crack even after $1.7 \times 10^5$ cycles. This is attributed to the crack growth resistance of SIC[76,77]. Moreover, the temperature change decreases from 4.2 K to 3.7 K (degradation degree of 12%) (Fig. 8 (a)) while the stress decreases from 1.3 MPa to 0.7 MPa (degradation degree of 46%) (Fig. 8 (b)) after $1.7 \times 10^5$ cycles. Finally, both temperature change and stress are stabilized. It primarily proves the feasibility of NR applicable for high-cycle cooling application. In conclusion, when an appropriate pre-strain is applied, to make the eC effect work at SIC strain regime, both a larger eC strength and longer fatigue life can be obtained. The direct and indirect measurements of eC effect of NR were also compared[54]. The indirect method of measuring the isothermal entropy change of eC effect of NR is a unique method since the direct measurement by using differential scanning calorimetry (DSC) seems to be impossible.

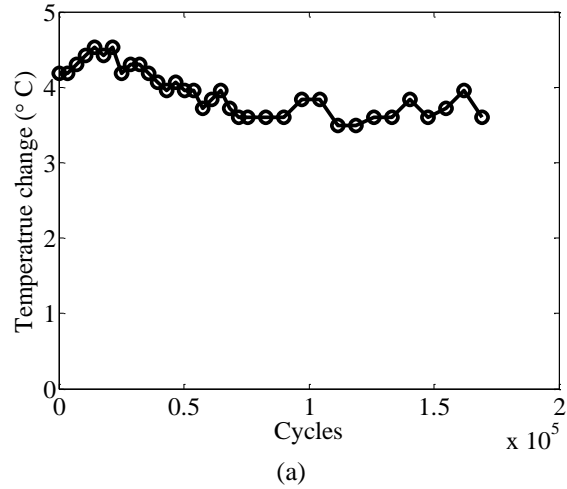

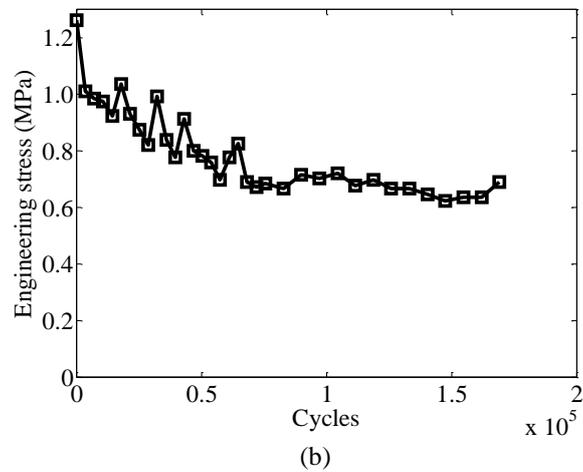

Fig. 8 The degradations of (a) eC temperature change ($\Delta T$) and (b) stress during $1.7 \times 10^5$ cycles at SIC strain regime (strain of 2-5). There is no crack, and a stable $\Delta T$ and stress are obtained finally (in press).

### 5.2.3 Potential study on elastocaloric effect of natural rubber based on SIC

The temperature dependence of eC effect of NR can be shown by the temperature dependence of SIC behavior. For NR, its glassy temperature is -70 °C and melting point is 220 °C, which can ensure the elastic behavior in the temperature range of cooling system (near room temperature). In the stretched state, its bearing temperature would decrease. Treloar[78] observed that the elasticity exists from -50 °C to 100 °C. Toki[79] observed that the stress began to decrease at 88 °C at constant strains, which may indicate the degradation of internal structure of NR. In conclusion, the requirement of the elasticity of NR near room temperature is satisfied.

For the temperature dependence of dynamic stretching, Rault *et al.*[80] measured the mechanical behavior and crystallinity with strain for a filled NR at different static temperatures (Fig. 9). They showed the existence of elasticity from 10 °C to 66 °C (Fig. 9 (a)), even though the maximum stress decreased with the increased temperature. The existence of crystallinity from 10 °C to 66 °C may indicate the existence of eC temperature change ($\Delta T$) in this near room temperature range. The decrease of crystallinity as temperature increases may predict the same decrease of $\Delta T$ with temperature. Direct measurement for temperature dependence of $\Delta T$ needs to be further conducted.

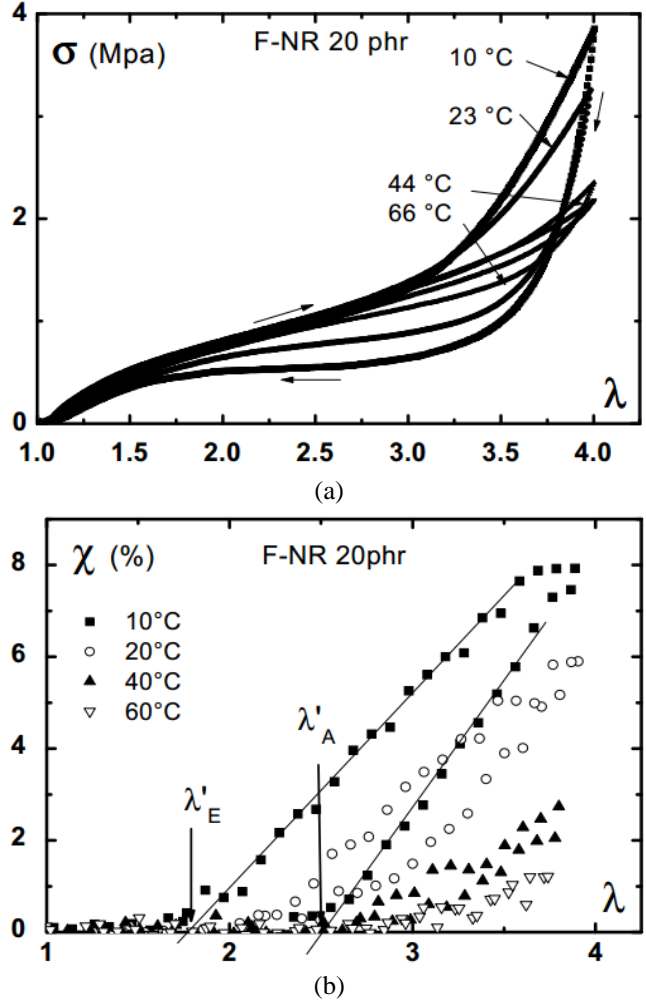

Fig. 9 Influence of temperature on: (a) stress-elongation and (b) crystallinity-elongation for a filled vulcanized NR[80].

As discussed above, SIC is responsible for both temperature change and fatigue property of eC effect of NR. The largest crystallinity occurs in the NR with an optimal network chain density ($1.2 \times 10^4$ mol/cm$^3$)[81], which may lead to the largest temperature change and the best fatigue property. Moreover, the filler (carbon black) can influence the crystallinity and the fatigue property of NR[82,83], and may thus influence its eC effect. To choose an optimal NR for eC effect, the direct influences of the network chain density and filler on the eC property of NR (temperature change and fatigue property) need to be further studied.

High-cycle fatigue test ($10^7$ cycles) towards application is needed. Moreover, the cooling device using NR material has not yet been investigated. A similar AMR system by using NR material should be conducted.

## 6. Applicable temperature change of caloric effect

For caloric materials, most of the reported $\Delta T$ and $\Delta S$ are the maximum values upon the ultimate external field (or close to) in a small number of cycles (for example, under the largest electric field before the EC material breaks down or under the largest deformation before the eC material breaks). The exploitation for larger temperature change upon an ultimate applied field would be over-studied. For eC effect, both the NR and SMAs show a short fatigue life ($10^3$-$10^4$ cycles) upon ultimate deformation cycle[84–86], which cannot be used for a cooling device ($10^7$ cycles is needed[13]).

Accordingly, these reported maximum ΔT are not applicable for a cooling device. The applicable ΔT should be measured from the applied field that allows enduring a large number of cycles ($10^7$ cycles). Decreasing strain/stress amplitude can be used for improving fatigue life of both NR[87,88] and SMAs[89].

For the eC effect of SMA, the fatigue life of $Ti_{54}Ni_{34}Cu_{12}$ can be up to $10^7$ cycles at strain amplitude of 1.5%[26]. The applicable temperature change should correspond to this strain for a cooling device. For a similar material, $Ti_{54.9}Ni_{32.5}Cu_{12.6}$, its temperature change is around 5 K upon strain of 1.5%[90]. Thus, the applicable ΔT for $Ti_{54}Ni_{34}Cu_{12}$ may be around 5 K. For the eC effect of NR, the fatigue life can be up to $10^7$ at strain amplitude of 200%[86]. The applicable temperature change corresponding to this strain amplitude may be up to 8 K[66]. This estimation is from another different NR material. The direct measurement of the applicable ΔT for eC effect after large number of cycles is needed and should be highlighted.

For the study of caloric material close to application, the applicable ΔT or ΔS should be an important evaluating factor. For EC effect, the applicable temperature change upon an electric field that allows enduring a large number of cycles has not been investigated.

## 7. Strategy for multicaloric effect

For the single-stimulus caloric effects, some caloric materials with high performance are not environmentally friendly, whereas for some environmentally friendly materials, they have low caloric performance. This problem may be solved by the multicaloric effect.

About the multi-caloric effect[38–40,42,43], it mainly focuses on two types: the coupling of electric field and mechanical field (uniaxial stress or hydrostatic pressure), and the coupling of magnetic field and mechanical field. Each coupling between the two stimulus fields can have three cases: (1) static stimulus 1 and dynamic stimulus 2; (2) dynamic stimulus 1 and static stimulus 2; (3) both stimulus 1 and 2 are dynamic. For the first and the second cases, they are mainly used to adjust the Curie temperature. For the conventional caloric effect, the stimulus will increase the Curie temperature. For the inverse caloric effect, the stimulus will reduce the Curie temperature[24]. EC effect and MC effect are generally tuned by the mechanical field. In turn, the BC / eC effect may also be tuned by electric/magnetic field. For the third case, while changing the Curie temperature, it can also be used to increase ΔS and ΔT. If both the caloric effects are the same type (both are conventional or inverse), in-phase stimulus can be applied to increase ΔS and ΔT. If they are opposite types of caloric effects, the anti-phase stimulus needs to be applied to increase ΔS and ΔT.

Practically, the EC effect is mainly coupled with a static mechanical field, whereas MC effect is coupled with both static and dynamic mechanical stimulus[38–40]. For the multi-caloric effect of EC effect, the application of the dynamic mechanical field will cause the crack of the electrode. When the electric field is applied, electric arc occurs in the crack and burns the material. Thus, to realize the application of the dynamic mechanical field and electric field simultaneously, the electrode should be compliant with the material[91]. However for the multi-caloric effect of MC effect, mechanical field can be easily applied dynamically.

In view of engineering, the application of multi-field is quite difficult and increases the cost. Moreover, the needed mechanical field can be up to the order of GPa[38,39], which in itself is a main drawback for mC effect of hard materials (more details in section 4 ). If a static mechanical field can be used to enhance the EC and MC effect, it should be easy to be applied and makes multi-caloric effect more attractive.

# 8. Comparison of elastocaloric effect of natural rubber with other caloric materials

## 8.1 Comparison of elastocaloric effect of natural rubber with shape memory alloys

Based on the temperature change $\Delta T$ of 12 K of NR, its isothermal entropy change $\Delta s$ can be estimated to be 80 $J \cdot K^{-1} \cdot kg^{-1}$ (70 $kJ \cdot K^{-1} \cdot m^{-3}$) by using $\Delta S = c \cdot \Delta T / T_0$, where $c$ is specific heat of NR. The SMAs exhibit $\Delta T$ ~20 K and $\Delta S$ ~37 $J \cdot K^{-1} \cdot kg^{-1}$ (240 $kJ \cdot K^{-1} \cdot m^{-3}$), as mentioned previously[14,15]. Both the $\Delta T$ and $\Delta S$ of NR and SMAs are larger than most of MC and EC materials (Fig. 10[92]). In the following part, the eC properties of NR and SMAs are compared.

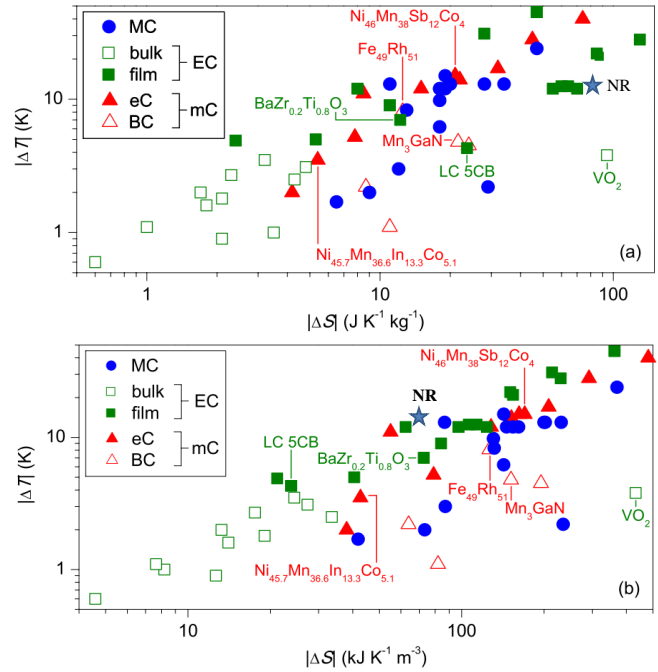

Fig. 10 Comparison of MC, EC and mC effects. Adiabatic temperature change |$\Delta T$| versus isothermal entropy change |$\Delta S$|, normalised by (a) mass and (b) volume. The normalization of $\Delta S$ by mass favors low-density EC polymers and eC NR, which are roughly four times less dense than EC ceramics and eC SMAs[92].

### 8.1.1 Comparison of mechanical behaviors of NR and SMAs

The tensile stress of SMAs can be up to several hundreds of MPa. Accordingly, the deformation of SMA is only several percent of its original length[30,93,94]. The small strain is an advantage of SMAs for a compact cooling device (small-scale application), whereas the large stress may be difficult to be provided (more details in section 4.1)[25].

For NR, the tensile stress is only several MPa, which is two orders of magnitude lower than SMAs (Table 1). It is because of the different phase: crystalline SMAs and amorphous NR. Accordingly, a larger eC strength of 5400 K/GPa can be obtained in NR, which is around two orders of magnitude larger than 13-45 K/GPa of SMAs (Table 1[24]). However, NR needs to be stretched several times of its original length[95], which is a drawback for the compact cooling device.

Table 1 Comparison of elastocaloric materials[24]

| Alloy system | Sample status | T(K) | |Δσ|(GPa) | |T|(K) | |T|/|Δσ|(K/GPa) |
|---|---|---|---|---|---|
| **$Cu_{68}Zn_{16}Al_{16}$** | Polycrystal | 328 | 0.275 | 6–7 | 21.8–25.5 |
| **$Ti_{50}Ni_{50}$** | Polycrystal | ~296 | 0.650 | 17 | 26.2 |
| **$Ti_{50.4}Ni_{49.6}$** | Thin film | 295 | 0.500 | 16 | 32 |
| **$Ti_{50.8}Ni_{49.2}$** | Ribbon | ~294 | ~0.46 | 14.1 | 30.7 |
| **$Ti_{54.9}Ni_{32.5}Cu_{12.6}$** | Thin film | ~346 | 0.350 | 6 | 17.1 |
| **$Ti_{55}Ni_{29.6}Cu_{12.6}Co_{2.8}$** | Thin film | 295 | 0.400 | 12 | 30 |
| **$Ni_{45.7}Mn_{36.6}In_{13.3}Co_{05.1}$** | [001]A polycrystal | 300 | 0.100 | 3.5 | 35 |
| **$Ni_{48}Mn_{35}In_{17}$** | [110]A polycrystal | ~313 | 0.300 | 4 | 13.3 |
| **$Ni_{50}Fe_{19}Ga_{27}Co_4$** | [001]A single crystal | 348 | 0.300 | 10 | 33.3 |
| **$Ni_{54}Fe_{19}Ga_{27}$** | Polycrystal | 298 | 0.133 | 6 | 45 |
| **$Fe_{68.8}Pd_{31.2}$** | [001]A single crystal | 240 | 0.100 | 2 | 20 |
| **Natural rubber** | **Amorphous** | **298** | **0.0016** | **8.7** | **5400**[54] |

The large deformation of NR can be reduced by pre-strain[55]. In view of the engineering stretching, the pre-strained length of NR sample can be considered as the initial length for the further stretching. Thus, the further strain for the pre-strained NR should be relative to this pre-strained length. A relative strain is defined as

$$\Delta\varepsilon_r = \frac{\Delta l}{l_{pre}} = \frac{\Delta l}{l_0 \cdot (1+\varepsilon_{pre})} = \frac{\Delta\varepsilon_m}{1+\varepsilon_{pre}}$$

where $\Delta l$ is the further stretched length (between pre-strain and final strain), $l_0$ is the initial length without pre-strain, $l_{pre}$ is the pre-strained length, $\varepsilon_{pre}$ is the pre-strain, $\Delta\varepsilon_m = \varepsilon_{max} - \varepsilon_{pre}$ is the strain variation that the NR material really experiences (real strain).

The relative strain is the really needed strain in engineering stretching. For the NR without pre-strain, the relative strain is equal to its real strain. As a pre-strain is applied, the relative strain is lower than the real strain. Fig. 11 shows an example of the relation between relative strain and real strain. Considering a pre-strain $\varepsilon_{pre} = 2$, the pre-strained length is $3l_0$. A further real strain variation is $\Delta\varepsilon_m = 2$, i.e. the length of sample changes from $3l_0$ to $5l_0$. Thus, the relative strain is only $2l_0/3l_0 = 0.67$. The needed strain can be largely decreased by applying a pre-strain.

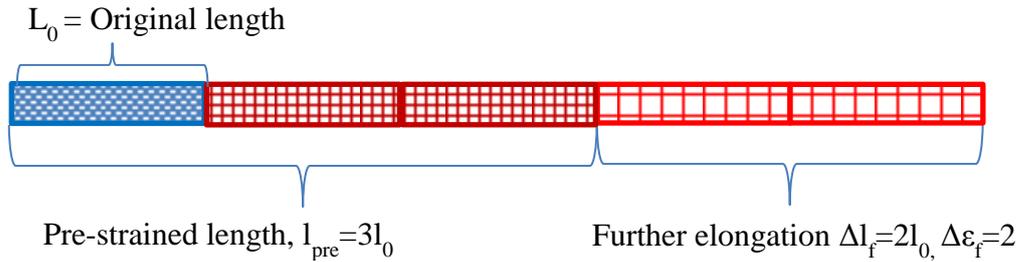

Fig. 11 The schematic diagram of the relative strain for the pre-strained sample.

Moreover, for current vapor-compression technology, the optimal compression volume ratio is 300%, which was found to result in the highest cooling density[96]. This value has been used in the SRM compressor K 608[97]. Thus, for the large-scale application, the large deformation in itself is not a problem.

### 8.1.2 Comparison of fatigue properties of NR and SMAs

For SMA, the latest paper reported that the fatigue life of $Ti_{54}Ni_{34}Cu_{12}$ in uniaxial stretching can be up to $10^7$ cycles at strain amplitude of 1.5%[26], which is about 1/4 of the ultimate strain amplitude. The fatigue property of NR may be better than SMAs. Cadwell et. al.[86] studied the dynamic fatigue life of NR upon a biaxial stretching (Fig. 12). They found that the fatigue life of NR depended on the minimum strain and the strain amplitude (ΔL). As the minimum strain increases, the fatigue life first increases and then decreases for the same strain amplitude. The longest fatigue life occurs when the minimum strain locates at middle strain (around 200%), which may correspond to the onset strain of SIC because SIC has a better crack growth resistance[76,77]. For strain amplitude, its decrease can increase the fatigue life. However, its decease can decrease the ΔT. Thus, a compromise should be made between the fatigue life and ΔT. Under the premise of satisfying the fatigue life for a cooling device，the strain amplitude should be chosen as large as possible for a larger ΔT. Middle strain amplitude of 200% (around 1/3 of the ultimate strain amplitude) with minimum strain of 200%, allows a fatigue life of $10^7$ cycles upon a biaxial stretching. The uniaxial stretching of NR should get a longer fatigue life. Thus, the fatigue requirement for a cooling device can be satisfied in NR. It should be mentioned that the influence of minimum strain/stress and strain amplitude has not been investigated in SMAs. Choosing an appropriate value for the two parameters may also improve the fatigue life of SMAs.

Compared with inorganic SMAs, the drawback of organic NR material is the aging problem, but its advantage is the healing mechanism. Normally, NR has a lifetime of 5-10 years and it may be reduced due to the stretching. Anti-degradant can be added to rubber compounds to partially avoid the deleterious effects of oxygen and ozone[98]. For the fatigue damage of NR, it can be healed by a high temperature or a solvent exposure[99]. Moreover, from the view of engineering, the cooling device can be designed with a replaceable part for NR material. This will be also beneficial from its low cost. This technique can also be analogized to other caloric materials with a poor fatigue property but low cost.

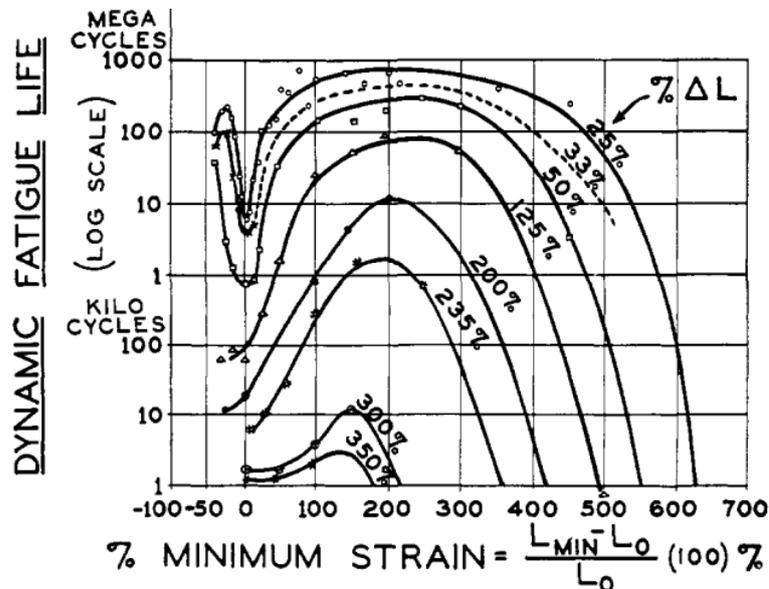

Fig. 12 The dynamic fatigue life of NR depends on minimum strain and strain amplitude[86].

Furthermore, NR is the most cost effective in all the giant caloric materials, which make it a competitive caloric material for commercialization. The comparison of SMAs and NR is summarized in Table 2.

Table 2 Comparison of the elastocaloric properties in hard materials and soft materials

|  | shape memory alloys (SMAs) | Natural rubber (NR) |
|---|---|---|
| **Cost** | 100 €/kg | 1.5 €/kg |
| **Fatigue life** | $10^7$ in uniaxial stretching (at strain amplitude of 1.5%) for $Ti_{54}Ni_{34}Cu_{12}$[26] | $10^7$ in biaxial stretching (at strain amplitude of 200%)[86] |
| **Stress ($\sigma$)** | hundreds of MPa; | several MPa; |
| **Strain ($\varepsilon$)** | several percent; | several hundreds of percent; |
| **Thermal conductivity** | 5.5 W.m$^{-1}$.K$^{-1}$[100] | 0.13 W.m$^{-1}$.K$^{-1}$ |
| **Adiabatic temperature change ($\Delta T$)** | ~ 20 K[14,15] | ~12 K[66] |
| **Isothermal entropy change ($\Delta s$)** | ~37 J.K$^{-1}$.kg$^{-1}$ (NiTi)[14,15] ~240 kJ.K$^{-1}$.m$^{-3}$ | ~80 J.K$^{-1}$.kg$^{-1}$[66] ~70 kJ.K$^{-1}$.m$^{-3}$ |

In conclusion, compared to SMAs as refrigerant, NR is low costly, requires less critical stress and has larger eC strength $\Delta T/\Delta\sigma$. However, it needs larger strain, has lower temperature change (but enough), less thermal conductivity and aging problem. The large strain of NR can be reduced by pre-strain (see section 8.1.1). For the aging problem, a replaceable part of NR refrigerant can be designed for the cooling device. For the fatigue life, both NR and SMAs can meet the fatigue requirement of a cooling device by choosing appropriate strain amplitude.

Due to the different eC properties of SMAs and NR, they may be adapted to different cooling application cases. Considering the large stress and high cost of SMAs, the SMAs film with a small force and low cost may be used and for small-scale application (see section 4.1). Considering the small tensile stress and low cost of NR, the bulk NR material with small tensile force and low cost can be used. It is thus potential for large-scale application because of the large cooling capacity and large volume.

### 8.2 Comparison of elastocaloric effect of NR with electrocaloric effect of PVDF-based polymers

The EC performance of PVDF-based polymers and eC effect of NR are compared in Table 3. There is no large difference for their caloric performance ($\Delta T$ or $\Delta S$). The remarkable differences are the cost and stimulus. For the cost, the NR is much lower than the PVDF-based polymers. For the stimulus, electric field is easier to be provided than the uniaxial stress (with a large strain) in small-scale system. Thus, the eC effect of NR material does not possess the advantage to EC effect of PVDF-based polymer for the small-scale application; even it has the advantage of low cost. However, for the large-scale application, where the large strain is not a problem, the PVDF-based polymer is incomparable with the NR material in terms of cost. As a result, both the eC effect of NR and EC effect of PVDF-based polymers should find their own application cases due to their own advantage in caloric material and stimulus.

Table 3 Comparison of caloric performance of the PVDF-based polymers and natural rubber (NR)

|  | PVDF-based polymers | NR |
|---|---|---|
| **Cost** | 2 €/g | 1.5 €/kg |
| **Stimulus** | Electric field or electric displacement | Uniaxial stress or strain |
| **Adiabatic temperature change ($\Delta T$)** | ~ 12 K[22] | ~12 K[66] |
| **Isothermal entropy change ($\Delta s$)** | ~55 J.K$^{-1}$.kg$^{-1}$[22] ~90 kJ.K$^{-1}$.m$^{-3}$ | ~80 J.K$^{-1}$.kg$^{-1}$[66] ~70 kJ.K$^{-1}$.m$^{-3}$ |

## Conclusion and perspectives

In the framework of solid-state cooling technology, four caloric effects, the magnetocaloric (MC), electrocaloric (EC), barocaloric (BC) and elastocaloric (eC) effects, are compared in view of environment and application. This field is mainly driven by the need of environmentally friendly technology. However, the caloric effect still shows the environmental problem. For the stimulus, the production of rare-earth elements (REEs) based permanent magnets for MC effect is detrimental to environment. For caloric materials, some mostly researched materials possess high caloric performance but they are toxic, like the arsenic (As)-based MC materials and lead-based EC materials. For the environmentally friendly caloric materials, they show a low caloric performance. The multicaloric effect may be employed to solve this problem, but it may increase the systematic complexity and cost. Some materials are environmentally friendly and possess high caloric performance but need high cost, like the PVDF-based EC materials. For BC and eC materials, the widely researched shape memory alloys (SMAs) is difficult to be manipulated due to their large stress. Thus, new caloric materials, which are environmentally friendly, cost-effective, of high caloric performance and practicable, need to be found.

### Evaluation for natural rubber

The eC effect of natural rubber (NR) is highlighted for cooling application in four aspects. From the environmental aspect, NR is environmentally friendly, non-toxic and recyclable. From the aspect of caloric performance, NR has a large temperature change ($\Delta T$) of 12 K, a large isothermal entropy change of 80 $J \cdot K^{-1} \cdot kg^{-1}$, and a long fatigue life of $10^7$ deformation cycles. From the practicable aspect, it needs a low tensile stress of several MPa, which is two orders of magnitude lower than SMAs. From the commercial aspect, its fabrication is a mature technology, resulting in the lowest cost among all the giant caloric materials. It is easily fabricated into the required shape for cooling device, like the tube.

The eC research of NR for cooling application is just beginning. The eC effect of NR is mainly related to the strain-induced crystallization (SIC). The SIC research can help to understand the eC effect of NR and to facilitate its application. For choosing deformation regime of eC effect, working in SIC strain regime leads to a larger $\Delta T$ and a longer fatigue life than working in the amorphous strain regime. A pre-strain method can be used to skip the amorphous strain regime and make the eC effect work directly in SIC strain regime. Moreover, the pre-strain can reduce the large strain of NR. Considering the working temperature span of NR material, the existence of SIC for a wide temperature span near room temperature may ensure a wide temperature span of $\Delta T$ near room temperature. For choosing NR material, it is proposed to choose the NR with optimal network chain density for a larger crystallinity, which may lead to a larger $\Delta T$ and a longer fatigue life. Thus, the research on eC effect of NR should consider fully the SIC theory. The oldest known caloric effect is the last to be exploited for cooling application, with plenty of scope for improvement and a bright future.

### Different caloric effects for different application cases

Currently, due to the large force needed for bulk SMAs, and the low dielectric strength of bulk EC material (ceramics and polymers), their film form is a better choice. Due to the small volume and small cooling capacity of the film, it is interesting for the small-scale cooling applications. Moreover, the SMAs, ceramics and polymers are expensive materials. The small quantity needed for small-scale application can make them commercially viable. This is a good correspondence of intrinsic property of caloric materials (small force for SMAs film and high performance for EC film) with the cooling application case (small volume, small cooling capacity and low cost of film needed for small-scale application). Furthermore, considering the stimulus, electric field is easier to provide than mechanical field in small-scale system, like the IT device. This is the absolute advantage of EC effect compared with other caloric effects and thus will decide its unique application case. Considering both property and stimulus of EC and eC material, EC effect is more appropriate for the small-scale cooling application.

For the large-scale application, further improvements of dielectric strength of the bulk EC materials are needed but for the low costly material. For SMAs, the decrease of force for bulk materials seems to be difficult. Elastomers, especially the natural rubber (NR), may be the best candidate for large-scale applications due to their low tensile stress. It has the same stress level as the current vapor-compressor, which proves its technical feasibility and facilitate its technology transformation. As for the large deformation of NR, it can be reduced by pre-strain on one hand. On the other hand, it faces the large-scale application and thus the large deformation in itself should not be a problem since its strain level is similar to that (volume ratio) of vapor-compressor. Moreover, the low cost of NR make the large quantity of NR commercially viable for large-scale application.

Cooling devices using NR have not yet been investigated. A similar cooling system to active magnetic regenerator (AMR) can be conducted. Considering the driving mode, the NR can be used to fabricate hand-driven cooling device due to its low tensile stress. Besides the application for refrigerator (cooling objects), the hand-driven cooling device can be used for cooling ourselves. It is different from the mode of cooling by means of improving heat transfer using a fan. It should be mentioned that, the hand-driven cooling device may not be realized by the SMAs due to their large stress.

**Acknowledgements**
The authors would like to thank China Scholarship Council (CSC). Thanks for the check from nonspecialist Dr. Yao Zhu.